\documentclass[journal]{IEEEtran}
\usepackage{graphicx}
\usepackage{cite}
\usepackage{amsmath}
\usepackage{amsfonts}
\usepackage{amssymb}
\usepackage{subfigure}
\usepackage{psfrag}
\usepackage{color}

\usepackage{tikz}

\newcommand{\bm}{\mathbf}
\newcommand{\be}{\begin{equation}}
\newcommand{\ee}{\end{equation}}
\newcommand{\bea}{\begin{eqnarray}}
\newcommand{\eea}{\end{eqnarray}}
\newcommand{\x}{{\bm x}}
\newcommand{\s}{{\bm s}}
\newcommand{\y}{{\bm y}}
\newcommand{\z}{{\bm z}}

\newcommand{\ba}{{\bm a}}
\newcommand{\br}{{\bm r}}

\newcommand{\dd}{{\bm d}}
\newcommand{\bA}{{\bm A}}

\newcommand{\bR}{{\bm R}}
\newcommand{\bW}{{\bm W}}

\newcommand{\bF}{{\bf F}}
\newcommand{\bG}{{\bf G}}

\newcommand{\bS}{{\bf S}}
\newcommand{\bH}{{\bf H}}
\newcommand{\bY}{{\bf Y}}
\newcommand{\bX}{{\bf X}}
\newcommand{\bV}{{\bf V}}
\newcommand{\bZ}{{\bf Z}}

\newcommand{\f}{{\bf f}}
\newcommand{\bzero}{{\bf 0}}

\newcommand{\eye}{{\bm I}}
\newcommand{\I}{{\bm I }}

\newcommand{\BW}{{\boldsymbol{\mathcal W}}}

\newcommand{\BH}{{\boldsymbol{\mathcal H}}}

\newcommand{\bnu}{\mbox{\boldmath$\nu$}}

\title{Low Complexity Modem Structure for OFDM-based Orthogonal Time Frequency Space Modulation}

\author{\normalsize Arman Farhang, Ahmad RezazadehReyhani, Linda E. Doyle, and Behrouz Farhang-Boroujeny
\vspace{-0.8 cm}
\thanks{
This publication has emanated from research conducted with the financial support of Science Foundation Ireland (SFI) and is co-funded under the European Regional Development Fund under Grant Number 13/RC/2077.

A. Farhang and L. E. Doyle are with Trinity College Dublin, Dublin 2, Ireland (e-mail: \{farhanga, ledoyle\}@tcd.ie).

A. RezazadehReyhani and B. Farhang-Boroujeny are with the University of Utah, USA (e-mail: \{Ahmad.Rezazadeh, farhang\}@ece.utah.edu).
}
}

\begin{document}

\maketitle

\begin{abstract}
Orthogonal time frequency space (OTFS) modulation is a two-dimensional signaling technique that has recently emerged in the literature to tackle the time-varying (TV) wireless channels. OTFS deploys the Doppler-delay plane to multiplex the transmit data where the time variations of the TV channel are integrated over time and hence the equivalent channel relating the input and output of the system boils down to a time-invariant one. This signaling technique can be implemented on the top of a given multicarrier waveform with the addition of precoding and post-processing units to the modulator and demodulator. In this paper, we present discrete-time formulation of an OFDM-based OTFS system. We argue against deployment of window functions at the OTFS transmitter in realistic scenarios and thus limit any sort of windowing to the receiver side. We study the channel impact in discrete-time providing deeper insights into OTFS systems. Moreover, our derivations lead to simplified modulator and demodulator structures that are far simpler than those in the literature.
\end{abstract}

\section{Introduction}\label{sec:Intro}
\IEEEPARstart{W}{hile} orthogonal frequency division multiplexing (OFDM) modulation scheme achieves a performance near the capacity limits in linear time-invariant channels, it leads to a poor performance in doubly dispersive channels. This is due to the large amount of interference that is imposed by the channel Doppler spread. The common approach to cope with this issue in the existing wireless standards such as IEEE 802.11a and digital video broadcasting systems is to shorten OFDM symbol duration in time so that the channel variations over each OFDM symbol are negligible. However, this reduces the spectral efficiency of transmission since the cyclic prefix (CP) length should remain constant. A thorough analysis of OFDM in such channels is conducted in \cite{Wang2006}. Another classical approach for handling the time-varying (TV) channels is to utilize filtered multicarrier systems that are optimized for a balanced performance in doubly dispersive channels, \cite{Farhang-Boroujeny2014}.  

Recently, new signaling techniques have emerged in the literature to tackle the TV channels, \cite{Dean2017,OTFS}. In \cite{Dean2017}, the authors introduce a new waveform called frequency-division multiplexing with a frequency-domain cyclic prefix (FDM-FDCP). This waveform, at its current stage, outperforms OFDM in channels with high Doppler and low delay spread. However, its performance in channels with high Doppler and medium to high delay spread is yet to be understood, \cite{Dean2017}. 

Orthogonal time frequency space (OTFS) modulation, which is the main focus of this paper, is another emerging signaling technique that is capable of handling the TV channels. OTFS was first introduced in the pioneering work of Handani et al., \cite{OTFS_WCNC}, where the two dimensional (2D) Doppler-delay domain was proposed for multiplexing the transmit data. OTFS modulation is a generalized signaling framework where precoding and post-processing units are added to the modulator and demodulator of a multicarrier waveform allowing for taking advantage of full time and frequency diversity gain of doubly dispersive channels. This process also coverts the TV channel to a time-invariant one. This modulation scheme has two stages. First, a set of complex data symbols in the Doppler-delay domain are converted to the time-frequency domain through an inverse symplectic finite Fourier transform (SFFT$^{-1}$), \cite{OTFS_WCNC}. In the second stage, the resulting time-frequency samples are fed into a multicarrier modulator to form the time domain transmit signal. The reverse operations are performed at the receiver to map the received signal back to the Doppler-delay domain. In addition, to further improve the channel sparsity in the Doppler-delay space, application of proper window functions to the 2D signals, after the SFFT$^{-1}$ at the transmitter and before the SFFT block at the receiver, is proposed in \cite{OTFS_WCNC}. However, very little is said on the choice of such windows. 
Moreover, although the OTFS formulations presented in \cite{OTFS_WCNC} are kept general, the numerical results presented there seem to be limited to the case where OFDM is used for transmission of the generated time-frequency signals.

In this paper, we present a discrete-time formulation of OTFS modulation while limiting ourselves to the case where OFDM is used for time-frequency signal modulation. This study reveals that in realistic scenarios this formulation leads to highly simplified transmitter and receiver structures compared with the existing ones in \cite{OTFS_WCNC,rakib2017orthogonal,rakib2017orthogonal_IoT,hadani2017compatible}. We note that window functions (at the transmitter and receiver) that may be effective in converting the channel to a sparse one in the Doppler-delay space require knowledge of channel variations on a given data packet at the transmitter which obviously is unknown. Hence, it is hard to say any effective window in this sense could be applied at the transmitter. On this basis, we will ignore the windowing step of the OTFS at the transmitter side. At the receiver side, on the other hand, an iterative channel estimation and equalization may allow the use of an effective window that leads to a sparse channel, e.g. \cite{Raviteja2017,Li2017}. This is an interesting topic that falls out of the scope of this paper and may be left for future studies. In this paper, we include the windowing of the time-frequency signal at the receiver side, but limit our discussion to the case where this is a separable window along the time and frequency dimensions and emphasise on channel independent window functions with some reasonable impact.  
\begin{figure*}
\psfrag{x}{\hspace{-1 mm}\small $x_{k,l}$}
\psfrag{y}{\hspace{2 mm}\small $y_{m,n}$}
\psfrag{z}{\hspace{2 mm}\small $z_{m,n}$}
\psfrag{ +}{\hspace{6.5 mm}\scriptsize $+$}
\psfrag{xhat}{\small $\tilde{x}_{k,l}$}
\psfrag{Ch}{\small LTV Channel}
\psfrag{Window}{\hspace{3.5 mm}\small SFFT}
\psfrag{SFFTinv}{\hspace{1 mm} \small SFFT$^{-1}$}
\psfrag{Tx}{\hspace{2 mm}\small ~OFDM-based OTFS Transmitter}
\psfrag{Rx}{\hspace{3.5 mm}\small ~OFDM-based OTFS Receiver}
\psfrag{OFDM_Tx}{\small\hspace{-0.1cm}OFDM}
\psfrag{OFDM}{\hspace{-3 mm}\small OFDM}
\psfrag{Demod}{\hspace{-6 mm}\small Demodulator}
\psfrag{Mod}{\hspace{-5.5 mm}\small Modulator}
\psfrag{SFFT}{\hspace{0.2 mm}\small Windowing }
\centering
\includegraphics[scale=0.28]{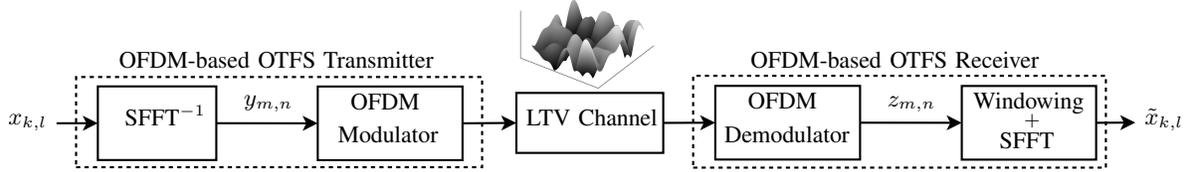} 
\vspace{-2 mm}
\caption{OTFS transmitter and receiver structure.}
\vspace{-5 mm}
\label{fig:OTFS}
\end{figure*}

The rest of this paper is organized as follows. In Section~\ref{sec:OTFS_Mod}, we present OFDM-based OTFS system model in its general form. Section~\ref{sec:Channel} presents the channel impact in the Doppler-delay domain. In Section~\ref{sec:Proposed_TxRx}, we present our proposed OFDM-based OTFS modem structure. We analyze the computational complexity of our proposed structure in Section~\ref{sec:Complexity} and finally, conclude the paper in Section~\ref{sec:Conclusions}.

\textit{Notations:} Throughout the paper, matrices, vectors and scalar quantities are denoted by boldface uppercase, boldface lowercase and normal letters, respectively. $[\bA]_{mn}$, and $\bA^{-1}$ represent the $mn^{\rm{th}}$ element, and the inverse of $\bA$, respectively. The function ${\rm vec}\{\bA\}$ vectorizes $\bA$ by stacking its columns on the top of one another in a column vector. The functions ${\rm circ}\{\ba\}$, and ${\rm diag}\{\ba\}$ form a circulant matrix whose first column is the vector $\ba$ and a diagonal matrix with the diagonal elements in $\ba$, respectively. $\I_m$ and $\bzero_{m\times n}$ are the identity matrix of size $m$ and an all zero $m\times n$ matrix, respectively. The superscripts $(\cdot)^{\rm T}$, $(\cdot)^{\rm H}$ and $(\cdot)^\ast$ indicate transpose, conjugate transpose and conjugate operations, respectively. Finally, the operators $\odot$, $\otimes$, ${\circledast}$, and $((\cdot))_N$ represent elementwise multiplication, Kronecker product, 2D circular convolution, and modulo-N operations, respectively.

\section{OFDM-based OTFS System}\label{sec:OTFS_Mod}
We consider an OTFS system transmitting a block of $M\times N$ Doppler-delay quadrature amplitude modulated (QAM) data symbols $x_{k,l}$ where $k = 0,\ldots,M-1$ and $l = 0,\ldots,N-1$. In OTFS modulation, the data symbols, $x_{k,l}$, are first converted to the time-frequency domain through the SFFT$^{-1}$ operation, 
\be\label{eqn:SFFTinv}
y_{m,n} =\sum_{k=0}^{M-1}{\sum_{l=0}^{N-1}{x_{k,l}b_{k,l}(m,n)}},
\ee
where $b_{k,l}(m,n) = \frac{1}{\sqrt{MN}}e^{-j2\pi(\frac{mk}{M}-\frac{nl}{N})}$. Due to the discussion in Section~\ref{sec:Intro}, we ignore transmit windowing step of OTFS. Hence, the time-frequency samples $y_{m,n}$ are directly fed into the OFDM transmitter to form the OTFS transmit signal,
\be\label{eqn:OTFS_St}
\bS = \bA_{\rm CP}\bF_M^{\rm H}\bY,
\ee
where $\bY$ is the $M\times N$ matrix containing the time-frequency samples $y_{m,n}$ on its $mn^{\rm th}$ elements. $\bF_M$ is the normalized $M$-point discrete Fourier transform (DFT) matrix with the elements $[\bF_M]_{pq}=\frac{1}{\sqrt{M}}e^{\frac{-j2\pi pq}{M}}$, for $p,q=0,\ldots,M-1$. $\bA_{\rm CP} = [\bG_{\rm CP}^{\rm T},\eye_M^{\rm T}]^{\rm T}$ is the cyclic prefix (CP) addition matrix where the $M_{\rm CP}\times M$ matrix $\bG_{\rm CP}$ is formed by taking the last $M_{\rm CP}$ rows of the identity matrix $\eye_M$ and $M_{\rm CP}$ is the CP length. Finally, the $(M+M_{\rm CP}) \times N$ matrix $\bS$ includes the OFDM time domain transmit signals on its columns. After parallel to serial conversion of $\bS$, the OTFS transmit signal at the baseband can be formed as $\s = {\rm vec}\{\bS \}$.

The received signal samples after transmission over a linear time varying (LTV) channel with the length $L$, $\{h(\kappa,\ell)\}_{\ell=0}^{L-1}$, can be obtained as 
\be\label{eqn:r_LTV}
r(\kappa) = \sum_{\ell=0}^{L-1}{h(\kappa,\ell)s(\kappa-\ell)+\nu(\kappa)},
\ee
where $\nu(\kappa)\sim\mathcal{CN}(0,\sigma_\nu^2)$ is the channel noise. Assuming $M_{\rm CP}\geq L$, the received OFDM symbols are free of intersymbol interference. Thus, the received OFDM symbol $n$ at the output of the OFDM demodulator can be written as
\be\label{eqn:y(n)bar}
{\z}_n =\bF_M{\bH}_n\bF_M^{\rm H}\y_{n}+\overline{\bnu}_n,
\ee
where ${\y}_n$ is the $n^{\rm th}$ column of $\bY$, ${\bH}_n=\bR_{\rm CP}\breve\bH_n\bA_{\rm CP}$, $\bR_{\rm CP}=[\bzero_{M\times M_{\rm CP}},\eye_M]$ is the CP removal matrix, and $\breve\bH_n$ is the LTV channel matrix realizing the linear convolution operation, (\ref{eqn:r_LTV}), on the samples of the $n^{\rm th}$ OFDM symbol. The vector $\overline{\bnu}_n$ contains the noise samples at the OFDM demodulator output and ${\z}_n=[z_{0,n},\ldots,z_{M-1,n}]^{\rm T}$. As the second step of the OTFS demodulator, the time-frequency samples, $z_{m,n}$, are windowed using a receive window function $w_{m,n}$, and then the result is converted back to the Doppler-delay domain through a SFFT operation. This process can be represented as
\be\label{eqn:xhat}
\tilde{x}_{k,l} =\sum_{m=0}^{M-1}{\sum_{n=0}^{N-1}{w_{m,n}z_{m,n}b_{k,l}^*(m,n)}},
\ee
where $k=0,\ldots,M-1$, and $l=0,\ldots,N-1$. It is shown in \cite{OTFS_WCNC} that the OTFS receiver output samples, $\tilde{x}_{k,l}$s, can be obtained as the 2D circular convolution of the QAM data symbols $x_{k,l}$ and a {\em time-invariant}   channel impulse response in the Doppler-delay domain. The OFDM-based OTFS modulation and demodulation process is shown in Fig.~\ref{fig:OTFS}. More details and derivations on the baseband channel response in the Doppler-delay domain are presented in the following section.

\section{Channel Impact}\label{sec:Channel}
To study the channel impact, in this section, we expand the formulation presented in Section~\ref{sec:OTFS_Mod}. In addition to providing a deeper insight into OTFS, this paves the way for the derivation of a novel modem structure in Section~\ref{sec:Proposed_TxRx}. The SFFT$^{-1}$ operation in (\ref{eqn:SFFTinv}) can be represented as 
\be\label{eqn:Y_TF}
\bY=\bF_M\bX\bF_N^{\rm H}, 
\ee
where $\bX$ is an $M\times N$ matrix containing the data symbols $x_{k,l}$ on its $kl^{\rm th}$ elements. From (\ref{eqn:Y_TF}), $\y_n=\bF_M\bX\f_{N,n}^*$ where $\f_{N,n}^*$ is the $n^{\rm th}$ column of the $N$-point DFT matrix. Consequently, (\ref{eqn:y(n)bar}) can be expanded as
\bea\label{eqn:y(n)bar_expanded}
\z_n &=&  \bF_M {\bH}_n\bF_M^{\rm H}\bF_M\bX\f_{N,n}^*+\overline{\bnu}_n \nonumber \\
&=& \bF_M {\bH}_n\bX\f_{N,n}^*+\overline{\bnu}_n.
\eea
Forming an $M\times N$ matrix ${\bZ}=[{\z}_0,\ldots,{\z}_{N-1}]$, the receiver windowing as well as the SFFT operation shown in (\ref{eqn:xhat}) can be rearranged as
\be\label{eqn:Xhat_w1}
\tilde{\bX}=\bF_M^{\rm H}(\bW\odot {\bZ})\bF_N,
\ee
where the $M\times N$ matrix $\tilde{\bX}$ contains the samples $\tilde{x}_{k,l}$ at the OTFS receiver output and the 2D window matrix $\bW$ includes $w_{m,n}$ on its $mn^{\rm th}$ elements. As mentioned in Section~\ref{sec:Intro}, here, we consider the general class of separable receiver window functions, i.e. $w_{m,n}=w^{\rm c}_m w^{\rm r}_n$, where the functions $w^{\rm c}_m$ and $w^{\rm r}_n$ window the columns and rows of ${\bZ}$, respectively. In this case, (\ref{eqn:Xhat_w1}) can be written as
\be\label{eqn:Xhat_w}
\tilde{\bX}=\bF_M^{\rm H}(\bW^{\rm c}{\bZ}\bW^{\rm r})\bF_N,
\ee
where $\bW^{\rm c}={\rm diag}\{[w^{\rm c}_0,\ldots,w^{\rm c}_{M-1}]\}$, and $\bW^{\rm r}={\rm diag}\{[w^{\rm r}_0,\ldots,w^{\rm r}_{N-1}]\}$.

To derive a clearcut representation of the channel effect on the transmit data symbols $x_{k,l}$s, we expand the $n^{\rm th}$ column of the matrix $\tilde{\bX}$ as
\bea\label{eqn:xtilden}
\tilde{\x}_n&=&\bF_M^{\rm H}(\bW^{\rm c}{\bZ}\bW^{\rm r})\f_{N,n} \nonumber \\
&=&\frac{1}{\sqrt{N}}\bF_M^{\rm H}\bW^{\rm c}\sum_{i=0}^{N-1}{{\z}_iw^{\rm r}_ie^{-\frac{j2\pi ni}{N}}},
\eea
Substituting ${\z}_i$ from (\ref{eqn:y(n)bar_expanded}) into (\ref{eqn:xtilden}), we have
\begin{align}\label{eqn:xhat(n)_final}
\tilde{\x}_n
&=\frac{1}{\sqrt{N}}\overline{\bW}^{\rm c}\sum_{i=0}^{N-1}{{\bH}_i\bX\f_{N,i}^*w^{\rm r}_i}e^{-\frac{j2\pi ni}{N}}+\tilde{\bnu}_n\nonumber\\
&= \frac{1}{N}\overline{\bW}^{\rm c}\sum_{i=0}^{N-1}\sum_{k=0}^{N-1}{{\bH}_i\x_kw^{\rm r}_i} e^{-\frac{j2\pi (n-k)i}{N}}+\tilde{\bnu}_n\nonumber \\
&= \overline{\bW}^{\rm c}\sum_{k=0}^{N-1}{\BH_{((n-k))_N}\x_k}+\tilde{\bnu}_n,
\end{align}
where $\tilde{\bnu}_n=\frac{1}{\sqrt{N}}\bF_M^{\rm H}\bW^{\rm c}\sum_{i=0}^{N-1}{\overline{\bnu}_iw^{\rm r}_ie^{-\frac{j2\pi ni}{N}}}$, $\overline{\bW}^{\rm c}=\bF_M^{\rm H}{\bW}^{\rm c}\bF_M$, and 
\be\label{eqn:BH_k}
\BH_{k}=\frac{1}{N}\sum_{i=0}^{N-1}{{\bH}_iw^{\rm r}_ie^{-\frac{j2\pi ki}{N}}}.
\ee
Next, defining an $MN\times MN$ block circulant matrix $\BH_{\rm BC}={\rm circ}\{[\BH_0^{\rm T},\ldots,\BH_{N-1}^{\rm T}]^{\rm T}\}$, the $MN\times MN$ block diagonal windowing matrix $\overline{\BW}^{\rm c} = \eye_N\otimes\overline{\bW}^{\rm c}$, the set of equations (\ref{eqn:xhat(n)_final}), for $n=0,1,\ldots,N-1$, can be combined together to obtain 
\be\label{eqn:dtilde}
\tilde{\dd}=(\overline{\BW}^{\rm c}\BH_{\rm BC})\dd+{\rm vec\{\tilde{\bV}\}},
\ee
where $\tilde{\dd}={\rm vec}\{\tilde{\bX}\}$, $\dd={\rm vec}\{\bX \}$, and $\tilde{\bV}=[\tilde{\bnu}_0,\ldots,\tilde{\bnu}_{N-1}]$. From (\ref{eqn:dtilde}), one may realize that if the submatrices $\BH_n$ in $\BH_{\rm BC}$ are circulant, i.e. the channel can be assumed time-invariant over each OFDM symbol, the multiplication of $\overline{\BW}^{\rm c}\BH_{\rm BC}$ to $\dd$ realizes the 2D circular convolution of the windowed Doppler-delay channel impulse response $\BH_{\rm DD,w}$ with the data matrix $\bX$. This is due to the fact that the matrices $\overline{\BW}^{\rm c}$, and $\BH_{\rm BC}$ are block circulant with circulant submatrices and their multiplication result in a matrix preserving their circulant property. Consequently, $\tilde{\bX}$ can be written as 
\be
\tilde{\bX}=\BH_{\rm DD,w} \circledast\bX+\tilde{\bV},
\ee
where the columns of the $M\times N$ matrix $\BH_{\rm DD,w}$ include the first columns of the matrices $\BH_n$ that are windowed by the window function $w^{\rm c}_m$ in the delay domain. In particular, the elements $kl$ of  $\BH_{\rm DD,w}$ are equal to $[ \overline{\bW}^{\rm c}\BH_l]_{k0}$ for $k = 0,\ldots,M-1$ and $l = 0,\ldots,N-1$. 

As a final note here, we note that (\ref{eqn:dtilde}) is a linear system of equations that relates $\tilde{\dd}$ to $\dd$. It, thus, may be used as a basis to derive linear detectors such as zero-forcing (ZF) and minimum mean square error (MMSE) ones or to develop soft detectors/equalizers.

\section{Proposed Modem Structure}\label{sec:Proposed_TxRx}
As discussed in the previous sections, our focus in this paper is on the general class of channel independent and separable window functions that are only applied at the receiver. In practical systems, without application of a window, the Doppler-delay channel impulse response along the delay dimension remains sparse and has the support of $L\ll M$. Any channel independent window along the frequency dimension, before application of the SFFT, translates into a circular convolution along the delay dimension. Most likely, this increases the support of the channel  impulse response, and hence, degrades the channel sparsity. In contrast, abrupt start and end of the subcarrier signals along the time dimension in the time-frequency signal matrix $\bZ$ translates to a non-sparse impulse response along the Doppler dimension. On these bases, we propose utilization of a rectangular window along the frequency domain, while using a window with smooth corners along the consecutive OFDM symbols in time.

\begin{figure}
\psfrag{yt}{\scriptsize$d_\kappa$}
\psfrag{M}{\hspace{-0.05cm}\scriptsize $M$}
\psfrag{z}{\hspace{-0.2cm}\scriptsize $Z^{-1}$}
\psfrag{PS}{\hspace{-0.1cm}\scriptsize P/S}
\psfrag{O}{\scriptsize $\s$}
\psfrag{P}{\hspace{-0.5cm} \scriptsize $N$-point}
\psfrag{D}{\hspace{-0.35cm} \scriptsize IDFT}
\psfrag{CP}{\hspace{-0.1cm}\scriptsize  CP}
\psfrag{ADD}{\hspace{-0.3cm}\scriptsize  Addition}
\centering
\includegraphics[scale=0.24]{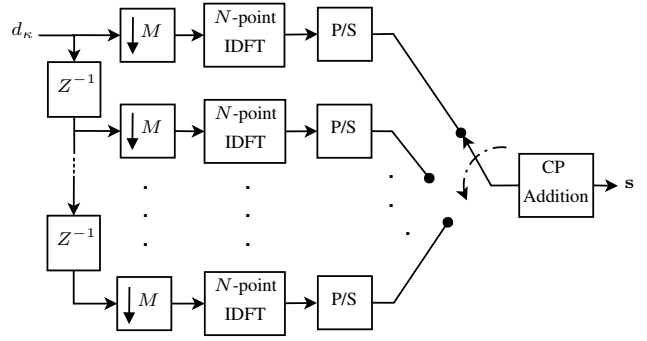} 
\vspace{-2 mm}
\caption{Our proposed OFDM-based OTFS modulator structure.}
\vspace{-3 mm}
\label{fig:OTFS_Tx}
\end{figure}

Next, we recall that the SFFT$^{-1}$ operation in (\ref{eqn:SFFTinv}) can be represented as $\bY=\bF_M\bX\bF_N^{\rm H}$. Inserting this into (\ref{eqn:OTFS_St}), one may realize that the IDFT operation at OFDM modulator and the $M$-point DFT operation in SFFT$^{-1}$ block cancel out. Consequently, (\ref{eqn:OTFS_St}) boils down to 
\be\label{eqn:S}
\bS = \bA_{\rm CP}\bX\bF_N^{\rm H}.
\ee
In (\ref{eqn:S}), the operation $\bX$ times $\bF_N^{\rm H}$ can be implemented through a set of M IDFT operations of size $N$ that are applied to the rows of $\bX$. CPs are then added to the columns of the resulting matrix. Fig.~\ref{fig:OTFS_Tx} presents a system structure that takes the QAM symbol elements of $\bX$ (denoted by the sequence $d_\kappa$), distributes them as inputs to the $M$ IDFT blocks, serializes and interleaves the results, and finally adds a CP prior to each block of $M$ samples at the output. The output here is the vectorized form of the matrix $\bS$.

\renewcommand{\arraystretch}{1.5}
\begin{table*}
  \centering
    \caption{Computational Complexity of Different Modem Structures}
    \label{tab:1}
{\begin{tabular}{|c|c|c|}
\hline\hline
Structure & Number of modulator CMs & Number of demodulator CMs  \\ \hline\hline
OTFS mod./demod. in \cite{OTFS_WCNC}  & $MN\log_2M+\frac{MN}{2}\log_2N$ & $MN\log_2M+\frac{MN}{2}(1+\log_2N)$ \\  \hline
OFDM mod./demod. & $\frac{MN}{2}\log_2M$ & $\frac{MN}{2}\log_2M$  \\ \hline
Our proposed mod./demod. & $\frac{MN}{2}\log_2N$ & $\frac{MN}{2}(1+\log_2N)$ \\ \hline\hline
    \end{tabular}}
\end{table*}

\begin{figure}
\psfrag{R}{\hspace{0cm}\scriptsize$\br$}
\psfrag{W}{\hspace{-0.5mm}\scriptsize$w^{\rm r}_{n}$}
\psfrag{M}{\hspace{-0.05cm}\scriptsize $M$}
\psfrag{z}{\hspace{-0.2cm}\scriptsize $Z^{-1}$}
\psfrag{PS}{\hspace{-0.1cm}\scriptsize P/S}
\psfrag{O}{\scriptsize $\tilde{d}_\kappa$}
\psfrag{P}{\hspace{-0.5cm} \scriptsize $N$-point}
\psfrag{D}{\hspace{-0.35cm} \scriptsize DFT}
\psfrag{CP}{\hspace{-0.1cm}\scriptsize  CP}
\psfrag{RMV}{\hspace{-0.3cm}\scriptsize  Removal}
\centering
\includegraphics[scale=0.24]{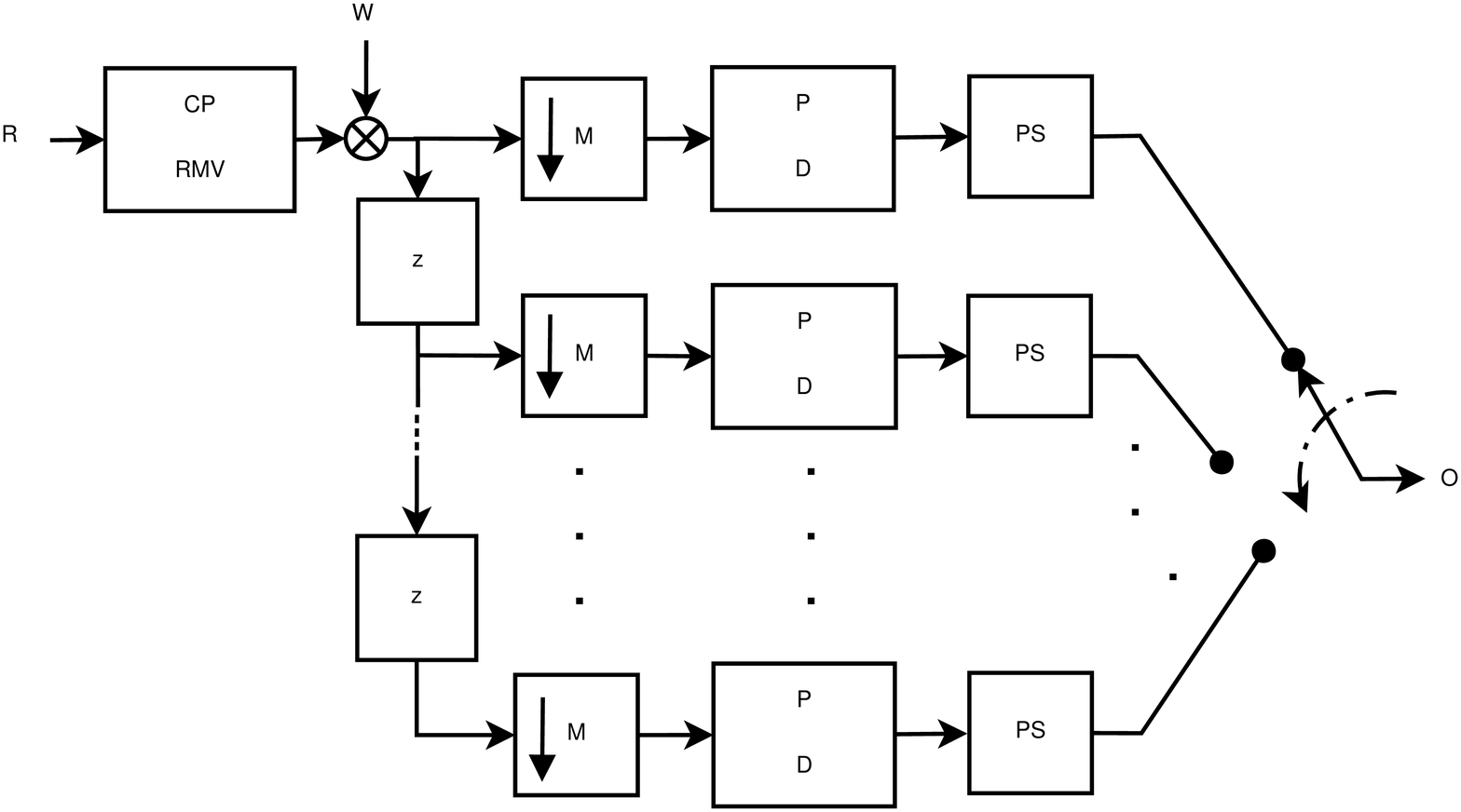} 
\vspace{-2 mm}
\caption{Our proposed OFDM-based OTFS demodulator structure.}
\vspace{-4 mm}
\label{fig:OTFS_Rx}
\end{figure}

To derive our proposed OTFS demodulator structure, we expand ${\bZ}$ as 
\begin{align}\label{eqn:Y1}
{\bZ}&=[\bF_M{\bH}_0\bX\f_{N,0}^*,\ldots,\bF_M{\bH}_{N-1}\bX\f_{N,N-1}^*]+\overline{\bV} \nonumber \\
&=\bF_M[{\bH}_0\bX\f_{N,0}^*,\ldots,{\bH}_{N-1}\bX\f_{N,0}^*]+\overline{\bV},
\end{align}
where $\overline{\bV}=[\overline{\bnu}_0,\ldots,\overline{\bnu}_{N-1}]$. Replacing $\overline{\bW}^{\rm c}$ with $\eye_M$ in (\ref{eqn:Xhat_w}), while using (\ref{eqn:Y1}), one realizes that similar to the transmitter, the DFT operation at the OFDM demodulator and the IDFT in the SFFT block cancel out. As a result, (\ref{eqn:Xhat_w}) can be expanded as
\be\label{eqn:xhat1}
\tilde{\bX}\!=\![w^{\rm r}_{0}{\bH}_0\bX\f_{N,0}^*,\ldots,w^{\rm r}_{N-1}{\bH}_{N-1}\bX\f_{N,N-1}^*]\bF_N+\tilde{\bV},
\ee
where the $M\times 1$ vectors $w^{\rm r}_{n}{\bH}_n\bX\f_{N,n}^*$ are the OFDM received symbols after the CP removal and application of the window coefficients $w^{\rm r}_{n}$. Also, $\tilde{\bV}=\bF_M^{\rm H}\bar{\bV}\bF_N$. Equation (\ref{eqn:xhat1}) shows that after CP removal, OTFS demodulator only requires scaling the OFDM symbols with the scalar $w^{\rm r}_{n}$ and application of DFTs (of size $N$) to the rows of the result. This leads to the OTFS demodulator structure that is presented in Fig.~\ref{fig:OTFS_Rx}. Here, the output sequence $\tilde{d}_\kappa$ is the serialized elements of $\tilde{\bX}$.

\section{Complexity Analysis}\label{sec:Complexity}
This section focuses on the computational complexity of the modulator and demodulator structures that were presented above and compares them with their counterparts in \cite{OTFS_WCNC} as well as those of OFDM.
Table.~\ref{tab:1} presents a summary of the computational complexity of different modulator/demodulator structures in terms of the respective number of complex multiplications (CMs). Clearly, the original OTFS structures that have been presented in \cite{OTFS_WCNC} and elsewhere, \cite{rakib2017orthogonal,rakib2017orthogonal_IoT,hadani2017compatible}, are significantly more complex than their OFDM counterparts. This is because of the separation of the inverse SFFT and SFFT blocks from the modulator and demodulator blocks, respectively. The OTFS structure proposed in this paper substantially reduces the complexity by combining the inverse SFFT and SFFT blocks with the OFDM modulator and demodulator blocks, respectively. Moreover, noting that in practical systems the number of OFDM symbols, $N$, in each data packet is much smaller than the FFT size, $M$, one will observe that our proposed OTFS modulator and demodulator blocks may have a significantly lower complexity than their OFDM counterparts.

\section{Conclusion}\label{sec:Conclusions}
In this paper, we studied OFDM-based OTFS modulation scheme. We presented the discrete-time formulation of such systems while investigating the channel impact on the transmit data symbols. An interesting finding of our study is that the SFFT$^{-1}$/SFFT and OFDM modulator/demodulator blocks in OTFS systems can be combined. This led us to a design of low complexity modulator and demodulator structures. Our complexity analysis revealed that our proposed OTFS modem structure offers a great amount of savings in computational complexity compared to the existing structures.

\bibliographystyle{IEEEtran} 

\end{document}